\documentclass[12pt,letterpaper]{article}
\global\def\draftcontrol{0}

\catcode`\@=11

\expandafter\ifx\csname draftcontrol\endcsname\relax\global\def\draftcontrol{0}
\fi

{\count255=\time\divide\count255 by 60
\xdef\hourmin{\number\count255}
\multiply\count255 by-60\advance\count255 by\time
\xdef\hourmin{\hourmin:\ifnum\count255<10 0\fi\the\count255}}
\def\draftdate{\number\month/\number\day/\number\year\ \ \ \hourmin }

\newcommand\makepapertitle{\par

  \begingroup
    \renewcommand\thefootnote{\@fnsymbol\c@footnote}%
    \def\@makefnmark{\rlap{\@textsuperscript{\normalfont\@thefnmark}}}%
    \long\def\@makefntext##1{\parindent 1em\noindent
            \hb@xt@1.8em{%
                \hss\@textsuperscript{\normalfont\@thefnmark}}##1}%
     \newpage
     \global\@topnum\z@   
     \@makepapertitle
     \thispagestyle{empty}\@thanks
  \endgroup
  \setcounter{footnote}{0}%
  \global\let\thanks\relax
  \global\let\makepapertitle\relax
  \global\let\@makepapertitle\relax
  \global\let\@thanks\@empty
  \global\let\@author\@empty
  \global\let\@date\@empty
  \global\let\@title\@empty
  \global\let\title\relax
  \global\let\author\relax
  \global\let\date\relax
  \global\let\and\relax
  \def\version{\let\version\@version\@gobble}
}
\def\@makepapertitle{%
  \newpage
   \ifnum\draftcontrol=1 {}
   \version\versionno
   \vskip 5em%
   \else
   \hfill\hbox to 3cm {\parbox{4cm}{\@pubnum}\hss}%
   \vskip 5em%
   \fi
   \begin{center}%
   \let \footnote \thanks
      {\hskip -0\textwidth \hbox to 1\textwidth%
        {\centerline{\Large\bf{\noindent\@title}}}}%
     \vskip 2em%
     {\normalsize
       \lineskip .5em%
       \begin{tabular}[t]{c}%
         \@author
       \end{tabular}\par}%
     \vskip 1em%
     {\@bstract}%
     \end{center}%
     \vfill
     \@date%
     \vskip 1.5em%
   \par
}

\gdef\@pubnum{}
\def\pubnum#1{%
  \gdef\@pubnum{#1}}

\gdef\@bstract{}
\def\Abstract#1{%
  \gdef\@bstract{%
   \parbox{\textwidth-0pc}{%
   \centerline{\bf Abstract}\penalty1000
   \noindent
   \renewcommand\baselinestretch{1.0}
   {#1}}}
}

\gdef\@email{}
\def\email#1{%
   \gdef\@email{%
   Email: {\tt #1}}
}

\def\ps@paper{\let\@mkboth\@gobbletwo%
     \ifnum\draftcontrol=1
      \def\@oddfoot{\hbox to \textwidth{\tiny \versionno \hfil\tiny\draftdate}%
        \hskip -\textwidth \hbox to \textwidth{\hfil\rm\thepage\hfil}}%
     \else\def\@oddfoot{\hbox to \textwidth{\hfil\rm\thepage\hfil}}
     \fi
     \let\@evenfoot\@oddfoot
}

\def\@version#1{\ifnum\draftcontrol=1
\typeout{}\typeout{#1}\typeout{}
\vskip3mm\centerline{\hbox{\fbox{\normalsize{\tt DRAFT -- #1 -- }
                   {\draftdate}}}}\vskip3mm
\fi}
\let\version\@version
\long\def\eqlabel#1{\ifnum\draftcontrol=1
                    \tag@false  
                    \tag*{(\theequation) \hbox to -0.2cm{\hspace{0cm}\small{#1}\hss}}
                    \refstepcounter{equation}
                    \edef\@currentlabel{\theequation}
                    \ltx@label{#1}          
                    \else
                    \label{#1}
                    \fi
                    }
\let\st@bibitem\@bibitem
\let\st@lbibitem\@lbibitem
\ifnum\draftcontrol=1
  \def\@bibitem#1{%
    \st@bibitem{#1}\a@@label{#1}\ignorespaces}
  \def\@lbibitem[#1]#2{%
    \st@lbibitem[#1]{#2}\a@@label{#2}\ignorespaces}
  \def\a@@label#1{%
    \gdef\a@lab{\smash{\normalfont\small#1}}
    \ifvmode
      \if@inlabel
        \global\setbox\@labels\hbox{%
          \llap{\a@lab\let\a@lab\relax
                \kern\@totalleftmargin\kern\marginparsep}%
          \box\@labels}%
      \fi
    \fi}
\fi

\usepackage{amsmath,bm,amsfonts,amssymb,array,calc,amsthm,rotating}
\usepackage{epsfig}
\usepackage[nosort]{cite}
\usepackage{graphicx}
\usepackage{color}
\usepackage[colorlinks=false]{hyperref}

\renewcommand\baselinestretch{1.25}
\renewcommand\section{\@startsection {section}{1}{\z@}%
                                   {-3.5ex \@plus -1ex \@minus -.2ex}%
                                   {2.3ex \@plus.2ex}%
                                   {\normalfont\normalsize\bfseries}}
\renewcommand\subsection{\@startsection{subsection}{2}{\z@}%
                                   {-3.25ex\@plus -1ex \@minus -.2ex}%
                                   {1.5ex \@plus .2ex}%
                                   {\normalfont\normalsize\it}}
\renewcommand\subsubsection{\@startsection{subsubsection}{3}{\z@}%
                                   {-3.25ex\@plus -1ex \@minus -.2ex}%
                                   {1.5ex \@plus .2ex}%
                                   {\normalfont\normalsize\it}}
\renewcommand\paragraph{\@startsection{paragraph}{4}{\z@}%
                                   {-3.25ex\@plus -1ex \@minus -.2ex}%
                                   {1.5ex \@plus .2ex}%
                                   {\normalfont\normalsize\bf}}
\renewcommand\subparagraph{\@startsection{subparagraph}{5}{\z@}%
                                   {-1.25ex\@plus -1ex \@minus -.2ex}%
                                   {0ex \@plus .2ex}%
                                   {\normalfont\normalsize\it}}

\numberwithin{equation}{section}

\long\def\@makecaption#1#2{%
  \vskip\abovecaptionskip
  \sbox\@tempboxa{{\bf #1:} #2}%
  \ifdim \wd\@tempboxa >\hsize
    {\small\bf #1:} {\small #2}\par
  \else
    \global \@minipagefalse
    \hb@xt@\hsize{\hfil\box\@tempboxa\hfil}%
  \fi
  \vskip\belowcaptionskip}

\renewcommand*\l@section[2]{%
  \ifnum \c@tocdepth >\z@
    \addpenalty\@secpenalty
    \addvspace{.5em \@plus\p@}%
    \setlength\@tempdima{1.5em}%
    \begingroup
      \parindent \z@ \rightskip \@pnumwidth
      \parfillskip -\@pnumwidth
      \leavevmode \bfseries
      \advance\leftskip\@tempdima
      \hskip -\leftskip
      #1\nobreak\hfil \nobreak\hb@xt@\@pnumwidth{\hss #2}\par
    \endgroup
  \fi}
\renewcommand*\l@subsection{\addvspace{.0em \@plus\p@}\@dottedtocline{2}{1.5em}{2.3em}}
\renewcommand*\l@subsubsection{\addvspace{-.2em \@plus\p@}\@dottedtocline{3}{3.8em}{3.2em}}

\def\o+{\oplus}

\def\beqa{\begin{eqnarray}}
\def\eeqa{\end{eqnarray}}

\newcommand{\al}{\alpha}

\newcommand{\si}{\sigma}

\newcommand{\cO}{{\cal O}}

\newcommand{\cV}{{\cal V}}

\def\Si{\Sigma}

\newcommand\iso{\kern.35em{\raise3pt\hbox{$\sim$}\kern -1.1em\to} \kern.3em}

\renewcommand {\thesection}{\arabic{section}}
\renewcommand{\theequation}{\thesection.\arabic{equation}}
\def\baselinestretch{1.3}

\begin{document}
\thispagestyle{empty}
\rightline{June 2007}

\vspace{2truecm}
\centerline{\LARGE Deformations of Bundles and the Standard Model}

\vspace{1.5truecm}
\centerline{Bj\"orn Andreas\footnote{supported by 
DFG-SFB 647 (G. C. is also supported by the MPI f\"ur Physik, M\"unchen)}$^2$ and Gottfried Curio$^1$$^3$}

\vspace{.6truecm}
\centerline{$^2${\em Institut f\"ur Mathematik, 
Freie Universit\"at Berlin}}
\centerline{\em Arnimallee 14, 14195 Berlin, Germany}
\centerline{$^3${\em Institut f\"ur Mathematik, 
Humboldt-Universit\"at zu Berlin}}
\centerline{{\em Rudower Chaussee 25, 12489 Berlin, Germany}}

\bigskip

\bigskip\bigskip
\noindent
We modify a recently proposed heterotic model hep-th/0703210, giving 
three net-generations of standard model fermions, to get 
rid of an additional $U(1)$ factor in the gauge group. 
The method employs a stable $SU(5)$ bundle on a Calabi-Yau 
threefold admitting a free involution. The bundle has to 
be built as a deformation of the direct sum 
of a stable $SU(4)$ bundle and the trivial line bundle. 
\newpage

In this note, which constitutes an addendum to \cite{AndCu07}, we propose a method to construct a model of the $E_8\times E_8$ heterotic string giving in four dimensions the gauge group and 
chiral matter content of the standard model. For this we embed an $SU(5)$ bundle 
in the first $E_8$ leading to a GUT gauge group $SU(5)$, which is afterwards broken by a 
Wilson line to the standard model gauge group. The non-simply connected Calabi-Yau threefold 
is obtained by modding a simply connected cover Calabi-Yau space $X$ 
by a free involution. Therefore we search on $X$ 
for an invariant $SU(5)$ bundle of net-generation number $\pm 6$ (for other constructions along these lines cf. \cite{Don00}, \cite{DoBu},
\cite{Bouchard:2006dn} ).

Models of this kind were recently constructed in \cite{AndCu07} 
(cf.\ also \cite{Blumenhagen:2005ga})
but had an additional $U(1)$ in the unbroken 
gauge group due to the specific form of the bundle
\begin{equation}
V_5=V_4\otimes \cO_X(-\pi^*\beta)\oplus \cO_X(4\pi^*\beta)
\label{fen}
\end{equation}
This is a polystable bundle and has structure group $SU(4)\times U(1)_A$ 
(on the Lie algebra level) of $V_5$ and therefore $SU(5)\times U(1)_A$ 
as unbroken gauge group. 

All the conditions, stability of the bundle, invariance 
under the involution, 
solution of the anomaly cancellation equation by having an effective 
five-brane class
and finally the phenomenologically net-generation number were therefore 
essentially 
solved already on the level of $V_4$. The bundle $V_4$ alone 
would give an unbroken
gauge group $SO(10)$, which cannot be broken to the standard model gauge group 
by just turning on a ${\bf Z}_2$ Wilson line corresponding to 
$\pi_1(X/{\bf Z}_2)$. Therefore 
in \cite{AndCu07} $V_4$ had to be enhanced to an $SU(5)$ bundle 
by adding a line bundle (the combined 
conditions of stability and five-brane effectivity make a non-trivial 
extension for $V_5$ problematical 
as explored in \cite{AndCu07}). Then the structure (\ref{fen}) 
caused the additional $U(1)_A$ in the gauge group.

 So in a $(4+1)$-decomposition of the rank $5$ structure group
one has 
$\left( \begin{array}{c|c} a&0\\ 
\hline
0&d\end{array}\right)\in SU(5)$ where 
the $U(1)_A$ is embedded as 
$\tiny{\left( \begin{array}{cccc|c} 1&&&&\\ &1&&&\\ &&1&&\\ &&&1&\\ 
\hline &&&& -4 \end{array}\right)}$. Our goal is to turn on 
the off-diagonal block elements to get a full, irreducible $SU(5)$, i.e.
$\left( \begin{array}{c|c} a&*\\ \hline *&d\end{array}\right)$. 
The possibility to do this will be measured (in first order) 
be two $Ext^1$-groups, corresponding to each of 
the off-diagonal blocks, respectively. The process of turning on these
off-diogonal terms means that the bundle $V'=V_4\oplus \cO_X$ 
(for simplicity we set $\beta=0$) is deformed to
a more generic bundle $V$; conversely $V'$ occurs as a specialization or
degeneration $V\to V'$ where the off-diagonal terms again go to zero and
one gets the reducible object.
More precisely, we say a vector bundle $V'$ deforms to a stable bundle 
$V$ if there is a connected curve $C$ and a vector bundle 
$\cV$ over $C\times X$ such that $V'\cong \cV_{\{0\}\times X}$ 
for some point $0\in C$ 
and $\cV_{\{t\}\times X}\cong V$
for some other point $t\in C$ 
\begin{eqnarray}
\cV \;\;\;\;\;\;\; V \;\;\;\;\;\longrightarrow \;\;\;\;\;\; V'\; \nonumber \\ \\
C \;\;\; \begin{array}{c|cc|c} &&&\nonumber \\ 
\hline t&\;\;\;\;\;\;\;\;\;\;\;\;\;\;\;& 0 & \end{array}
\end{eqnarray}

This process will thereby cause the following changes in
the structure group $G$ of the bundles and the unbroken 
gauge group $H$ of the four-dimensional low-energy observer
who sees the commutator in $E_8$ of $G$
\begin{equation}
\begin{array}{c|ccc} &V& \; \longrightarrow \; & V' \\ 
\hline
G&SU(5)&&SU(4)\times U(1)_A\\
H&SU(5)&&SU(5)\times U(1)_A
\end{array}
\end{equation}
For a generic choice of parameters with $\alpha \beta \neq 0$
(where $\alpha\in H^{11}(B)$ is a further twist class inherent 
in the construction, cf. \cite{AndCu07} and below) one would 
find that the $U(1)_A$, which occurs in the structure group 
and in the gauge group, is anomalous and thereby gets massive 
by the Green-Schwarz mechanism. However the group theory of the
$ad\, E_8$ decomposition relevant here tells
us that to secure the absence of exotic matter multiplets one has 
just to impose the condition $\alpha \beta = 0$ (cf.\cite{AndCu07}).
Thereby the $U(1)_A$ remains non-anomalous and remains in the light 
spectrum. It is this problem for which the present paper shows a way
out by embedding the reducible bundle in a family of proper irreducible $SU(5)$
bundles where the $U(1)_A$ is therefore again massive on the 
compactification scale (as all other elements of $E_8$ 
which are broken for the four-dimensional low-energy observer 
by the specific gauge background turned on on $X$); 
for the specialization $*\longrightarrow 0$,
where the off-diagonal elements are turned off again, one would then
get a restauration of four-dimensional gauge symmetry as this corresponds to
$m_{U(1)_A}\longrightarrow 0$, i.e., at this special point
on the boundary of the bundle moduli space the $U(1)_A$ 
returns into the light spectrum.

So, to get rid of this additional $U(1)_A$ factor we will construct 
a stable holomorphic 
$SU(5)$ bundle $V_5$ by deforming the complex structure of the given 
polystable $SU(5)$ bundle $V'=V_4\oplus \cO_X$ (as said, for simplicity we work 
from now on with $\beta=0$). $V_4$ is a stable $SU(4)$ bundle, and $\cO_X$ 
is the trivial one-dimensional line bundle; therefore $V'$ is a polystable 
bundle and solves the Donaldson-Uhlenbeck-Yau (DUY) equations. 
If such a deformation to a stable holomorphic bundle exist, the theorems 
of \cite{UY}, \cite{Don} guarantee that $V_5$ is a solution of the DUY 
equations, i.e., the equations of motion of the heterotic string. 

In \cite{Huyb} (Corollary B.3) it has been shown that the direct sum 
of two stable vector bundles (say $V, W$) of the same slope $\mu(V)=\int c_1(V)J^2/rk(V)$ deforms to 
a stable vector bundle if the sum has unobstructed deformations and both 
spaces $H^1(X, Hom(V, W))$ and $H^1(X, Hom(W,V))$ do not vanish. 
Applied to our case we therefore have to show that $H^1(X, Hom(V_4,\cO_X))$ 
and $H^1(X,Hom(\cO_X, V_4))$ do not vanish and that $V_4\oplus \cO_X$ 
has unobstructed deformations. 

$H^1(X, End(V'))$ is the space of all first-order 
deformations
of $V'$. The obstruction to extending a first order deformation to second 
(or higher) order lives in $H^2(X, End_0(V'))$. 
Thus if $H^2(X, End_0(V'))=0$ we can always lift to higher order, 
i.e., the deformations would be unobstructed. For instance, 
the tangent bundle $TX$ of a $K3$ surface has unobstructed deformations 
since $H^2(X, End_0(TX))\cong H^0(X, End_0(TX))^*=0$. 
As we are on a Calabi-Yau threefold, Serre duality 
shows that the dimensions of $H^1(X, End(V'))$ and $H^2(X, End(V'))$ 
are equal, so there are as many obstructions as deformations. 

The vanishing of $H^2(X, End(V'))$ is, however, only a sufficient condition 
for the existence of a global (in contrast to first-order) deformation.
So in principle it is still possible to have global deformations 
although the obstruction space is non-vanishing. That this hope 
is not in vain is born out by the example of $X$ being the quintic 
in ${\bf P}^4$ and $V'=TX\oplus \cO_X$ (this example was considered 
first in \cite{wit86}, \cite{wilou87}), where nevertheless knows that a 
global deformation exists \cite{Huyb}, \cite{dry}.  

The tangent space $H^1(X, {End}(V'))$ 
to the deformations decomposes as follows (using the fact 
that $H^1(X, \cO_X)=0$) 
\begin{equation}
H^1(X, End(V'))\cong H^1(X, End(V_4))\oplus 
H^1(X, Hom(V_4,\cO_X))\oplus H^1(X,Hom(\cO_X, V_4))\nonumber
\end{equation}
where the last two terms $Ext^1(V_4,\cO_X)=H^1(X,V_4^*)$ and 
$Ext^1(\cO_X,V_4)=H^1(X,V_4)$
parametrize non-trivial extensions
\begin{eqnarray}
0&\to& \cO_X\to W\to V_4\to 0\\
0&\to& V_4\to W'\to\cO_X\to 0
\end{eqnarray}
We note first that the index theorem gives (by stability of $V_4$ 
and $\mu(V_4)=0$ we have 
$H^i(X, V_4)=0$ for $i=0,3$; note that also $V^*$ is stable and 
has $\mu(V^*)=0$) 
\begin{equation}
{\rm dim}H^1(X,V_4)-{\rm dim}H^1(X, V_4^*)=-\frac{1}{2}c_3(V_4)
\end{equation}
which for the physical relevant $V_4$ is non-zero, so at least 
one of the two off-diagonal spaces is
already non-vanishing. To actually prove that $H^1(X,V_4)$ 
and $H^1(X, V_4^*)$ are both non-vanishing we recall first the explicit 
construction of $V_4$ form \cite{AndCu07}. Note that whereas 
in \cite{AndCu07} we considered the case $x>0$, $\beta\neq 0$, we consider 
here the case $x<0$ (this simplifies some arguments below) and $\beta=0$ . 

As in \cite{AndCu07} the rank four vector bundle $V_4$ will be 
constructed as an extension 
\begin{equation}
0\to \pi^*E_1\otimes \cO_X(-D)\to V_4\to \pi^*E_2\otimes \cO_X(D)\to0
\end{equation}
where $E_i$ are stable bundles on $B$ and $D=x\Si+\pi^*\al$ is a 
divisor in $X$ (here $X$ is a Calabi-Yau threefold elliptically fibered 
with two sections $\si_i$ over $B={\bf P}^1\times {\bf P}^1$,
cf. \cite{AndCu07}; furthermore $\Si=\si_1+\si_2$ and $F$ will denote the fiber). 
The argument for stability of $V_4$ runs exactly parallel to the one 
given in \cite{AndCu07}. To prove stability of 
$V_4$ we first note that given zero slope stable vector bundles $E_i$ on $B$, 
one can prove that the
pullback bundles $\pi^*E_i$ are stable on $X$ \cite{AndCu06}, \cite{AndCu06phys} for a suitable K\"ahler class $J$.
Now for the zero slope bundle $V_4$ constructed as an extension 
(with $\pi^*E_i$ stable) we have two immediate conditions which 
are necessary for stability: first that $\mu(\pi^*E_1\otimes \cO_X(-D))<0$ 
and second that $\pi^*E_2\otimes \cO_X(D)$ of 
$\mu(\pi^*E_2\otimes \cO_X(D))>0$ is not a subbundle of $V_4$, 
i.e., the extension is non-split. The first condition reduces to 
\begin{equation}
DJ^2=2x(h-z)^2c_1^2+2z(2h-z)\al c_1>0
\label{dj}
\end{equation}
(where $c_i:=\pi^*c_i(B)$) this implies in our case $x<0$ the condition
 \begin{equation}
 \al c_1>0
 \end{equation}
The non-split condition can be expressed as 
$Ext^1(\pi^*E_2\otimes \cO_X(D), \pi^*E_1\otimes 
\cO_X(-D))=H^1(X, {\cal E}\otimes \cO_X(-2D))\neq 0$ 
where ${\cal E}=E_1\otimes E_2^*$. As in \cite{AndCu07} 
applying the Leray spectral sequence to $\pi\colon X\to B$ 
yields as sufficient condition for $H^1(X, {\cal E}\otimes \cO_X(-2D))\neq 0$ 
the following condition (for $x<0$)
\begin{equation}
\chi(B, {\cal E}\otimes \cO_B(-2\al))=4+8\al^2-4\al c_1-2(k_1+k_2)<0
\label{ind}
\end{equation}
Finally, it remains to determine the range in the K\"ahler cone where 
$V_4$ is stable, i.e., that for any coherent subsheaf $F$ of rank $0<r<4$ 
we have $\mu(F)<0$.
Solving as in \cite{AndCu07} the corresponding inequalities we find for 
the general K\"ahler class  $J=z\Si+h\pi^*c_1$ where $z,h\in {\bf R}$ with $0<z<h$ the range 
(for $x<0$; here $\zeta:=h-z$)
\begin{equation}
\frac{-xc_1^2}{(\al-xc_1)c_1}h^2< h^2-\zeta^2<
\frac{-xc_1^2}{(\al-xc_1)c_1-1}h^2
\label{win}
\end{equation}
In summary, we find $V_4$ is stable if (\ref{ind}) and 
(\ref{win}) are satisfied (the latter just fixes an appropriate range of $z$). 

Let us now derive sufficient conditions for $H^1(X, V_4)\neq 0$ 
and $H^1(X, V_4^*)\neq 0$. The Leray spectral sequence gives the following exact sequence
\begin{equation}
0\to H^1(B,\pi_*V_4^*)\to H^1(X,V_4^*)\to 
\end{equation}
Thus it suffices to show that $H^1(B, \pi_*V_4^*)\neq 0$. For this we
apply $\pi_*$ to the defining exact sequence of $V_4^*$ and find the exact sequence
\begin{eqnarray}
0&\to &E_2^*\otimes \cO_B(-\al)\otimes \pi_*\cO_X(-x\Si)\to 
\pi_*V_4^*\to E_1^*\otimes \cO_B(\al)\otimes \pi_*\cO_X(x\Si)\to\nonumber
\end{eqnarray}
For $x<0$ one has $\pi_*\cO_X(x\Si)=0$ and finds 
$E_2^*\otimes\cO_B(-\al)\otimes \pi_*\cO_X(-x\Si)\cong\pi_*V_4^*$
It follows 
\begin{eqnarray}
H^1(B,\pi_*V_4^*)=H^1(B,E_2^*\otimes \cO_B(-\al)\otimes \pi_*\cO_X(-x\Si))
\end{eqnarray}
As $\pi_*\cO_X(-x\Si)=\cO_B\oplus\dots$ it will be sufficient to show 
that $H^1(B, E_2^*\otimes \cO_B(-\al))\neq 0$. The index theorem gives
$
\chi(B, E_2^*\otimes \cO_B(-\al))=2+\al^2-\al c_1-k_2
$
from which we conclude that 
\begin{equation}
2+\al^2-\al c_1-k_2<0 \;\Longrightarrow 
H^1(B, E_2^*\otimes \cO_B(-\al))\neq 0 \:\Longrightarrow H^1(X, V^*)\neq 0
\end{equation}
The same reasoning applied for $H^1(X, V)$ yields
\begin{equation}
2+\al^2-\al c_1-k_1<0 \;\Longrightarrow 
H^1(B, E_1^*\otimes \cO_B(-\al))\neq 0 \;\Longrightarrow H^1(X, V)\neq 0
\end{equation}
Let us now determine the physical constraints we have to impose on $V_5$ 
in order to get a 
viable standard model compactification of the heterotic string. 
What concerns the invariance of the deformed bundle one can, 
as in \cite{AndCu07} app.~B, argue for the existence of invariant 
elements in the two non-trivial extension spaces 
(to solve both conditions simultaneously one can use 
reflection twists $v\to -v$ in both fiber vector spaces of $V_4$ and $\cO_X$). Further one has still to make sure that the deformability 
to first order, which we have checked, extends to a full global construction, which we assume can be done.

Note further that the characteristic 
classes of $V'$ are invariant under deformations. Therefore we have 
$c(V_5)=c(V')$ and a direct computation yields
\begin{eqnarray}
c_2(V_5)&=&-2x(2\al-xc_1)\Si-2\al^2+k_1+k_2\\
\frac{c_3(V_5)}{2}&=&2x(k_1-k_2)
\end{eqnarray}
Further one has to satisfy the heterotic anomaly condition 
$c_2(X)-c_2(V_5)=[W]=w_B\Si+a_fF$ where $W$ is a space-time 
filling fivebrane wrapping a holomorphic curve of $X$. 
This leads to the condition 
that $[W]$ is an effective curve class in $X$, which in turn can be 
expressed by the two conditions $w_B\geq 0$ and $a_f\geq 0$. 
Inserting the expressions for $c_2(X)=6c_1\Si+5c_1^2+c_2$ and 
$c_2(V)$ gives the conditions
\begin{eqnarray}
w_B&=&(6-2x^2)c_1+4x\al\geq 0\\
a_f&=&44+2\al^2-k_1-k_2\geq 0
 \end{eqnarray}
Finally, the physical net-generation number of chiral fermions, 
downstairs on $X/{\bf Z}_2$, is given by
\begin{equation}
N_{gen}^{phys}=x(k_1-k_2)
\end{equation} 

To summarize, we get the following list of constraints (besides $x<0$)
\begin{eqnarray}
\al c_1&>& 0\\
2+\al^2-\al c_1-k_i&<&0, \ \ \ \ {\rm where}\ \ i=1,2\\
2+4\al^2-2\al c_1-(k_1+k_2)&<&0\\
\label{wb} (6-2x^2)c_1+4x\al&\geq &0\\
44+2\al^2-(k_1+k_2)&\geq &0\\
x(k_1-k_2)&=&\pm 3
\end{eqnarray}
(and $k_i\geq 8$ for $h=\frac{1}{2}$, cf. \cite{AndCu07}, app.~B). 
One realizes that (\ref{wb}) entails $x=-1$ and so
\begin{equation}
\al \leq c_1
\end{equation}
One finds that the following $\alpha$'s are possible 
(the entries in $(p,q)$ refer to the multiples of the 
two generators in $B={\bf P}^1\times {\bf P}^1$): 
\begin{equation}
\al=(-1,2),(1,1),(1,0),(0,2),(1,2),(2,2)
\end{equation}
besides interchanging the entries. For instance, one finds then for 
$\al=(1,1)$ 
that $k_1=8+i$, $k_2=11+i$ where $i=0,\dots, 14$ or $\al=(1,0)$ and 
$i=0,\dots,12$
(besides interchanging the $k_i$).

\end{document}